\documentclass{svjour2}              

\usepackage{array} 
\usepackage{amsmath}
\usepackage{amssymb}
\usepackage{latexsym}
\usepackage{color}
\usepackage{graphicx}

\newcommand{\sgn}[0]{\textrm{sgn}}

\newcommand{\RM}[1]{\MakeUppercase{\romannumeral #1{}}}
\begin{document}

\title{From hardcore Bosons to free Fermions with Painlev{\'e} V}


\author{Christian Recher  \and Heiner Kohler}


\institute{C. Recher  \& H. Kohler \at Instituto de Ciencias Materiales de Madrid, CSIC
              \\ Sor Juana In{\'e}s de la Cruz 3\\28049 Madrid,  Spain\\ \email{hkohler@icmm.csic.es} \\\\
            \emph{C. Recher \at Fakult\"at f\"ur Physik, Universit\"at Duisburg-Essen \\Lotharstrasse 1\\
        47048 Duisburg,     Germany \\ \email{christian.recher@uni-due.de}         }}

\date{Received: date / Accepted: date}

\maketitle

\begin{abstract}
We calculate   zero temperature Green's function, the density--density
correlation and expectation values of a one--dimensional quantum
particle which interacts with a Fermi--sea via a
$\delta$--potential. The  eigenfunctions of the Bethe-Ansatz
solvable model can be expressed as a determinant. This allows us
to obtain a compact expression for the Green's function of the extra particle.
 In the
hardcore limit the resulting expression can be analysed further
using Painlev{\'e} V transcendents. It is found that depending on
the extra particles momentum its Green's function undergoes a
transition of that for hardcore Bosons to that of free
Fermions. 

\keywords{Bethe-Ansatz\and exactly solvable models \and
Painlev{\'e} equations}
 \PACS{05.30 $\pm$ d \and 02.30.Ik \and 03.65.Yz}
\end{abstract}

\section{Introduction}
\label{intro}

Many intriguing questions are related to the dynamics  of
one--dimensional systems \cite{gia04,ima11}. Landau's
Fermi--liquid theory fails for Fermionic systems in one--dimension
and the Luttinger-liquid model  applies \cite{hal81}. The free Fermi--gas as a
fixed point of the renormalization group is unstable in favor of
collective Bosonic excitations. This makes it difficult for an
injected Fermionic particle to deposit its excess energy and
yields largely different electronic transport properties in
one--dimension \cite{gia04}. Corrections to the Luttinger--liquid
model were studied recently experimentally \cite{bar10} and
theoretically \cite{ima11}. On the other hand the unusual slow
decay of a one--dimensional interacting Boson system has been
observed \cite{kin06} and was related to the exact classical and
quantum solvability of the Lieb--Liniger model \cite{lie63,tak99}. This model was studied using Luttinger liquid 
theory in \cite{caz04}.

If in a 1d gas different species of particles are present the
theoretical description becomes more difficult, since the
many--body wave function transforms according to a higher
dimensional representation of the permutation group. This results
in additional quantum numbers apart from the asymptotically free
momenta, which have no classical analogue. Although for
$\delta$-interaction the eigenfunctions have been  constructed via the
nested Bethe--Ansatz \cite{yan67,gau66} the problem of calculating
Green's functions, respectively correlation functions are only
solved in some special cases \cite{kor93,che04b,ima06}.
In spin-$\frac{\scriptstyle 1}{\scriptstyle 2}$ systems the number
of additional quantum numbers is infinite in the thermodynamic
limit.  However, if one spin polarization, say spin up, is carried
by just one particle,  only one additional quantum number appears.
As we will show it can be identified with the free momentum of the
extra particle.
The eigenfunctions of this model as well as the extra particle's
density--density correlator were calculated by McGuire
\cite{mcg65,mcg66}. He focussed on the ground state, i.~e. the
additional quantum number is not present  in McGuire's work. 
If one is interested in the dynamical properties of the distinguishable
particle it is crucial to include it in order to describe
excitations of the extra particle. 

We consider McGuire's model where the distinguishable particle is
allowed to carry an arbitrary momentum and study its response to
the interaction with a Fermi--sea, which is assumed to be at zero
temperature.  It might be viewed upon as an extremely imbalanced
mixtures of two particle species. Their study was pioneered by Lai
and Yang \cite{lai71} and has revived boisterously in the recent
years both theoretically \cite{ima06,ima06a,gub08,yin08,mol98}  and
experimentally~\cite{tru01,pil09,ols09}.

We write the full Bethe--Ansatz many--body wave function as a
determinant and are thus able to derive compact determinantal
expressions for the extra particle's equal time Green's function
 (or likewise reduced density matrix). In the Tonks--Girardeau
regime of infinite interaction strength we employ the powerful
methods of T\"oplitz determinants and Painlev{\'e} equations,
which were applied earlier to hardcore Bosons
\cite{len72,wid73,jim80,for02,for03}. We find that in the hardcore
limit the particle's expectation values and its Green's function
depend crucially on its momentum. If the momentum is right at the
Fermi edge the extra particle behaves just like an additional
Fermion of the sea. The energy shift is zero and its single
particle Green's function is that of a free Fermion. If the
particle's momentum is in the core of the Fermi--sea, the energy
shift is finite and the Green's function is identical with that of
a hardcore Boson. As the particle's momentum  varies from the
Fermi edge to the core, the particle undergoes a transition from a
free Fermion to a hardcore Boson, manifested in an algebraic
asymptotic decay of the real space Green's function $\lim_{x\gg 1}
G(x)\sim x^{-\beta}$ with an exponent $\beta$ changing from one to
$1/2$. A small distance expansion of the Green's function and
numerical work show that for finite interaction strength this
picture does not change qualitatively.

The article is organized as follows: In Sec.~\ref{Model} we
introduce the model. Due to its importance in the present context,
we give a detailed discussion of the additional quantum number
appearing in the Bethe-Ansatz equations. Furthermore expectation
values are calculated. In Sec.~\ref{green} we derive a
determinantal expression for the Green's function and analyze it
further in the limit of small distances and in the hardcore limit. 
In Sec.~\ref{DD-corr} the density-density correlation function is
derived. We summarize and conclude in Sec.~\ref{summary}.

\section{Model}
\label{Model}

We consider a $N+1$ particle system  in one dimension where $N$
particles are identical spinless Fermions to which we refer as
environment or Fermi--sea. In addition we assume the presence of a
single particle which is distinguishable from the former ones. The
particles interact via a repulsive $\delta$--potential. Due to the
Pauli--principle, it acts only between the extra particle and the
particles of the Fermi-sea. The particle masses are all equal.
Choosing  the units such that $\hbar=1$ and $m=1/2$ the
Hamiltonian reads
\begin{eqnarray}
\hat{H}&=& -\sum_{j=1}^N\frac{\partial^2}{\partial
             x_j^2}- \frac{\partial^2}{\partial y^2} + 4c\sum_{j=1}^N
\delta(x_j -y) \label{eq1.1}.
\end{eqnarray}
Here the coordinates $x_j,~ j=1,\ldots,N$ denote the positions of
the $N$ identical Fermions while $y$ refers to the position of the
distinguishable particle. Furthermore $c\geq 0$ denotes the
interaction strength and the factor 4 is included for convenience.
The model is exactly solvable via  Bethe's--Ansatz
\cite{mcg65,mcg66}. However the original form of the wave function
constructed by this method is inconvenient for explicit
calculations.

\subsection{Eigenfunctions and Bethe--Ansatz Equations}

Denoting an eigenfunction of the Hamiltonian \eqref{eq1.1} by
$\Psi(x_1,\ldots,x_N,y)$, we require it to be antisymmetric in the
first $N$ arguments. This and the fact that there is no
interaction between the particles of the Fermi-sea suggest to write
$\Psi(x_1,\ldots,x_N,y)$ as a determinant
\begin{eqnarray}
\Psi(x_1,\ldots,x_N,y)&=& C_N \det\left[ A_j(x_l-y)e^{\imath k_j x_l}\Big|e^{ k_j
y}\right]_{\genfrac{}{}{0pt}{2}{j=1,\ldots,N+1}{
l=1,\ldots,N}}\label{eq1.1.2},\\
 A_j(x)&=& \imath(k_j -\Lambda) +c\, \sgn (x).\nonumber
\end{eqnarray}
Note that $\Psi(x_1,\ldots,x_N,y)$ has all required symmetries i.e. is
antisymmetric  under the exchange of all quasimomenta, antisymmetric
in the particle postions $x_l$ , $l=,1,\ldots,N$ and has no 
well defined symmetry when the distinguishable particle and a 
particle of the Fermi--sea are exchanged.
As shown in Sec.~\ref{sec3.2} the normalization constant $C_N$
is for periodic boundary conditions determined by (see Eq.~\eqref{eq4.7}) 
\begin{eqnarray}
\left|C_N\right|^{-2}=N!L \prod_{j=1}^{N+1}[ L((k_j-\Lambda)^2 +c^2 +2c/L)] \sum_{j=1}^{N+1}\frac{1}{ L((k_j-\Lambda)^2 +c^2 +2c/L) }.\nonumber\\\label{const}
\end{eqnarray}
In Eq.~\eqref{eq1.1.2} $\{k_j\}_{j=1,\ldots ,N+1}$ are the
"quasimomenta" and $\Lambda$ is an additional parameter (not
present in \cite{mcg65,mcg66}).

The wave function~\eqref{eq1.1.2} is an eigenfunction of the
Hamiltonian~\eqref{eq1.1} to the eigenvalue
\begin{eqnarray}
E=\sum_{j=1}^{N+1}k_j^2\label{eq1.1.2-1}.
\end{eqnarray}
Furthermore the eigenvalue to the center of mass momentum operator is
\begin{eqnarray}
K=\sum_{j=1}^{N+1} k_j\label{eq1.1.2-2}.
\end{eqnarray}
The form (\ref{eq1.1.2}) is crucial for our approach, since it
allows us to employ powerful methods of matrix algebra to
manipulate determinants.
Equations~\eqref{eq1.1.2}-\eqref{eq1.1.2-2} can be proved
straightforwardly by acting with $\hat{H}$ on $\Psi$.

In order to acquire results  for finite
particle density we impose  periodic boundary conditions
\begin{eqnarray}
\Psi(x_1,\ldots,x_l,\ldots,x_N,y)&=&\Psi(x_1,\ldots,x_l+L,\ldots,x_N,y)\ , \  l=1,\ldots,N ,\label{eq1.2.1}\\
\Psi(x_1,\ldots,x_N,y)&=&\Psi(x_1,\ldots,x_N,y+L)\label{eq1.2.2} \ .
\end{eqnarray}
The first condition yields the
transcendental equations
\begin{eqnarray}
 e^{\imath k_j L}  &=& \frac{i(k_j -\Lambda) +c \sgn(x-y)}{i(k_j -\Lambda) +c \sgn(x +L-y)} \ ,\quad j=1,\ldots,N+1\label{eq1.2.3}
\end{eqnarray}
which after taking the logarithm on both sides translates to
\begin{eqnarray}
 k_j L &=&2\pi n_j-2\arctan \left(\frac{k_j -\Lambda}{c}\right),\ \ \quad j=1,\ldots,N+1\ .\label{eq1.2.3-1}
 \end{eqnarray}
As can be seen by considering the limit $c \to 0$  of
Eq.~\eqref{eq1.2.3-1} the numbers $n_j$  have to be half-integers
and  $-\pi/2$ $<\arctan(x)$ $<\pi/2$. The second condition
\eqref{eq1.2.2} yields
\begin{eqnarray}
k_jL = 2\pi \tilde{n}_j +2\sum_{n\neq
j}^{N+1}\arctan\left(\frac{k_n -\Lambda}{c}\right),\qquad
j=1,\ldots,N+1\label{eq1.2.4},
\end{eqnarray}
where $\tilde{n}_j$ are half-integers for $N$ odd and integers for
$N$ even. In the following we will for convenience always assume
$N$ to be odd.

Combining Eq.~\eqref{eq1.2.3} and Eq.~\eqref{eq1.2.4}  yields
quantization rules for the quasimomenta $k_j$ and $\Lambda$  which
are known as Bethe--Ansatz equations
\begin{eqnarray}
k_jL&=& 2 \pi \left(m_j+ \frac{1}{2}\right) - 2 \arctan \left( \frac{k_j-\Lambda}{c}\right) \ , \ \  j=1,\ldots,N+1\label{eq1.2.5},\\
2\pi J_{\Lambda}&=&2\sum_{j=1}^{N+1}\arctan\left(\frac{k_j-\Lambda}{c}\right)\label{eq1.2.6}  \ .
\end{eqnarray}
Now the $m_j$   are unequal integers. According to
Eq.~\eqref{eq1.2.6} the integer $J_{\Lambda}$ is bounded by
\begin{eqnarray}
-\frac{N+1}{2} \leq J_{\Lambda} \leq \frac{N+1}{2}.\label{eq1.2.6-1}
\end{eqnarray}
The quantities $m_j ,J_{\Lambda} $   are the quantum numbers
labeling an eigenstate of the system. For $J_\Lambda=\Lambda$ $=0$
an expression equivalent to Eq.~\eqref{eq1.2.5}  was derived in
Ref.~\cite{mcg65}. For a derivation of Eqs.~\eqref{eq1.2.5} and
\eqref{eq1.2.6} from the Bethe--Ansatz wave function we refer to
Ref.~\cite{tak99}.

\subsection{Solution of Bethe--Ansatz Equations and Ground State}

The Bethe-Ansatz equations are coupled algebraic equations which
in general do not permit a closed solution. In our case the
situation is more favorable, since the equations for $k_j$ only
couple via $\Lambda$. For $c\to 0$ and for $c\to$ $\infty$ they
can be solved exactly.

Let $\{m_j\}_{j=1,\ldots,N+1}$ be a set of ordered integers
$m_j>m_{j+1}$. Then in the limit $c\to 0^+$ a solution to the
Bethe--Ansatz equations (\ref{eq1.2.5}) and (\ref{eq1.2.6}) is
given by
\begin{eqnarray}
\label{czero}
k_j & = & \left\{\begin{array}{ll}\displaystyle\frac{2\pi}{L} m_j^+\ ,&
 j \leq J_\Lambda+\frac{N+1}{2},\\[1em]
            \displaystyle \frac{2\pi}{L} \left(m_j+1\right)^- \ , &j > J_\Lambda+\frac{N+1}{2}\end{array}\right.
\end{eqnarray}
and $\Lambda \in [m_j,m_{j+1}+1]$. The superscript $\pm$ indicates
that $k_j$ is approaching its value from above/below. For $c\to
0^+$ the solution for $\Lambda$ is in general not unique. We come
back to this point below.

Since the quantum numbers  $m_j$ and $J_\Lambda$  for the ground
state  are the same for $c=0$ and $c>0$, the quantum numbers $m_j$
of the interacting ground state can be extracted from
Eq.~(\ref{czero})
\begin{eqnarray}\label{ms}
\{m_j\}_{j=1,\ldots,N+1}&=&\left\{\frac{N-1}{2} ,\ldots,0,\ldots ,-\frac{N+1}{2}\right\} \ .
\end{eqnarray}
This yields for $c=0$  the set of quasimomenta
\begin{eqnarray}
\{ k_j\}_{j=1,\ldots,N+1}&=&\frac{2\pi}{L}\left\{0^+,0^-,\pm 1,\ldots,\pm \frac{N-1}{2}\right\} \ .
\end{eqnarray}
Since the distinguishable particle can occupy the same state as a
sea-particle, in the ground state the single particle
quasimomentum with the lowest energy is double occupied. Thus for
the ground state  $J_\Lambda=0$. The allowed interval for
$\Lambda$ shrinks to a point and the  solution for $\Lambda$
becomes unique $\Lambda= 0$. The same happens for any other state
for which $m_{J_{\Lambda+(N+1)/2}}$ and
$m_{(J_{\Lambda+(N+1)/2}+1)}$ are adjacent integers.

In the hardcore limit  $c\rightarrow \infty$
Eqs.~\eqref{eq1.2.5} and \eqref{eq1.2.6} become
\begin{eqnarray}
k_jL&=&2\pi \left(m_j+\frac{1}{2}\right) + 2\arctan(\lambda), \label{eq1.2.9}\\
\lambda&=&-\tan\left(\frac{\pi J_{\Lambda}}{N+1}\right)\ , \quad \lambda \ =\ \frac{\Lambda}{ c}
\label{eq1.2.10},
\end{eqnarray}
where the dimensionless factor $\lambda$ takes into account that
$\Lambda$ scales like $c$ for $c\rightarrow \infty$.
From the definition in Eq.~\eqref{eq1.2.10} it is seen that
$\lambda$ can range between $-\infty$ and $\infty$ when
$J_{\Lambda}$ varies according to Eq.~\eqref{eq1.2.6-1}.

For finite $c$ the Bethe--Ansatz equations can be calculated
numerically.  As $c$ increases the  quasimomenta $k_j$ evolve
smoothly from $2\pi (m_j+\theta(\Lambda-m_j))/L$ to
$2\pi(m_j+{\scriptstyle \frac{1}{2}}\sgn(\Lambda-m_j))L$, i.~e.~from
multiple integer to multiple half--integer values of $2\pi/L$. The
additional parameter $\Lambda$ evolves smoothly from its initial
value to $\pm\infty$ as the interaction increases.

In the following we assume that the sea--particles are in the
ground state but the extra particle may occupy an arbitrary single
particle state.  Two cases have to be distinguished:

Either the quasimomentum of the extra particle lies inside the
Fermi--sea.  Then the quantum numbers $m_j$ of the state are given
by the set of integers in Eq.~(\ref{ms}). The non--interacting
extra particle shares its momentum with one of the particles of
the Fermi--sea.  The additional quantum number $J_\Lambda$
indicates the double occupied quasimomentum. For $J_\Lambda =
(N-1)/2$ or $J_\Lambda$ $=-(N-1)/2$ the Fermi--momentum of the sea
is double occupied. The non--interacting eigenfunctions for a
given set of quasimomenta $\{k_j\}_{j=1,\ldots,N+1}$
is uniquely determined and the energy--eigenvalue is
non--degenerate. Thus for $c=0$ the energy increases quadratically
with $J_\Lambda$.

\begin{figure}
\centering
\includegraphics{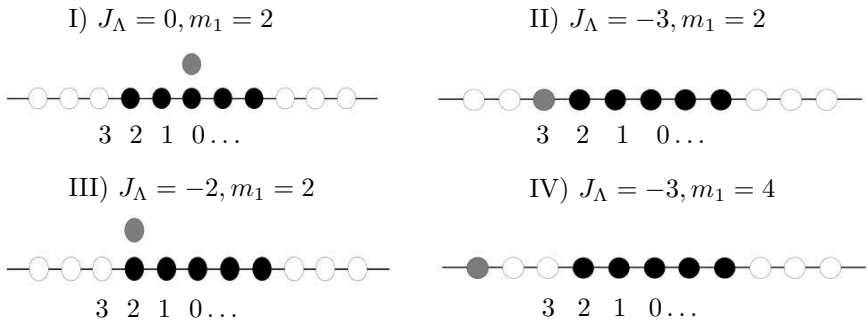}
\caption{\label{fig1} Sketch of the occupied non--interacting
one--particle states for $N=5$. A black circle denotes a state
occupied by a sea--Fermion, an empty circle denotes an empty
state. The grey circle denotes the state, which is occupied by the
extra particle. Case I) is the ground state configuration. In case
II) and III) the quantum numbers $m_j$ are given by
Eq.~\eqref{ms}. In case II) the extra particle's momentum is just
outside the Fermi--sea in case III) it is just inside.}
\end{figure}

In the second case  the quasimomentum of the extra particle lies
outside the Fermi--sea. The quantum number $J_\Lambda$ takes its
value at the upper edge $J_\Lambda = (N+1)/2$ or at the  lower
edge  $J_\Lambda = -(N+1)/2$ of its spectrum.  For vanishing
interaction all $N+1$ quasimomenta are different.  If the
highest/lowest quantum number is $m_1 = (N-1)/2$ respectively
$m_{N+1} = -(N+1)/2$ the extra particle's momentum borders the
Fermi--sea.  If $m_1>(N-1)/2$ or $m_{N+1}<-(N+1)/2$ the extra
particle's momentum is outside the Fermi sea.   For a given set of
quasimomenta  there exist $N+1$ orthogonal eigenfunctions since
the extra particle might carry any of the $k_n$'s without changing
the systems energy. These eigenfunctions are distinguished by
$J_\Lambda$. Thus at the Fermi--edge the energy becomes
independent of $J_\Lambda$ and the density of states has a
singularity. The different cases are illustrated in
Fig.~\ref{fig1}.

Summation over Eq.~\eqref{eq1.2.5} from $j=1$ to $j=N+1$ yields in
combination with Eq.~\eqref{eq1.2.6}  for the overall momentum of
the system
\begin{eqnarray}
K=\sum_{j=1}^{N+1}k_j= \frac{2\pi}{L}\left(\sum_{j=1}^{N+1} \left(m_j +\frac{1}{2}\right) - J_{\Lambda}\right)\label{eq1.2.6-2}.
\end{eqnarray}
If the set $\{m_j\}_{j=1,\ldots ,N+1}$ is given by Eq.~(\ref{ms})
the sum vanishes and the quantum number $J_\Lambda= -
\frac{LK}{2\pi}$ is directly related with the center of mass
momentum $K$. The latter in turn is identified with the momentum
of the extra particle in the lab frame where the Fermi--sea is at
rest.

\subsection{Thermodynamic Limit}
\label{thermlim}

For a dense set of quantum numbers as in Eq.~(\ref{ms}) the
density of states $\varrho (k)= \partial m(k) /\partial k$ can be
derived by taking the derivative of Eq.~\eqref{eq1.2.5} 
\begin{eqnarray}
\varrho(k)= \frac{L}{2\pi} + \frac{1}{\pi} \frac{c}{(k-\Lambda)^2 +c^2} \label{eq1.2.7}.
\end{eqnarray}
This expression has a rather natural interpretation: The first
term corresponds to the density of states of a non-interacting
Fermi--sea. The second term appears due to the presence of the
distinguishable particle. It has the form of a Lorentzian
distribution centered around $\Lambda$ with width $c$. For
$c\rightarrow 0$ it becomes  a $\delta$-function at $k=\Lambda$.

In the thermodynamic limit, where $N,L\rightarrow \infty$ such
that the particle density $\rho=N/L$ remains finite there are
different choices for the scaling of the interaction strength
\begin{equation}\label{cscaling}
c \ = \             \left\{\begin{array}{ll}\bar{c} L^{-1} \\[0.5em]
                                                          \hat{c}\left(\pi N/L\right)\\[0.5em]
                                                          \infty \quad & \mbox{\rm (harcore limit) \
                                                          }\end{array}. \right.
\end{equation}
The Lorentzian part in Eq.~\eqref{eq1.2.7} is outscaled by the
part due to the Fermi-sea in all three cases. This implies that in
the thermodynamic limit the single particle does not affect the
spectrum of the Fermi-sea. The effect of the particle onto the
bath should therefore be negligible. This is of course an
essential requirement for a {\em bona fide} dissipative quantum
system.

The system's total energy is
\begin{equation}
E\ = \ \int\limits_{-k_{\rm F}}^{+k_{\rm F}} dk k^2\varrho(k)\label{eq1.2.1.1} \ .
\end{equation}
For the energy shift due to the interaction this yields up to
corrections of order $1/N$
\begin{eqnarray}
\label{shift}
E- E_0  &=& k_{\rm F}^2 -\frac{k_{\rm{F}}^2-\Lambda^2 +c^2}{\pi} v(k_{\rm F},c,K)-\left(\frac{2\pi}{L}J_\Lambda\right)^2 \\
&& +\frac{c}{\pi}\left[2k_{\rm F} +\Lambda\ln\left(\frac{c^2+(\Lambda- k_{\rm F})^2}{c^2+(\Lambda+ k_{\rm F})^2}\right)\right],\nonumber
\end{eqnarray}
where the Fermi-momentum is $k_{\rm F} = \pi \rho+ {\cal O}(N^0)$ and
\begin{equation}
E_0 \ = \frac{ \pi^2 N^3}{3L^2} +\left(\frac{2\pi}{L}J_\Lambda\right)^2 +{\cal O}(N^0)
\end{equation}
corresponds to the energy of the non--interacting system.
In the thermodynamic limit where the
quasimomenta are distributed according to Eq.~\eqref{ms}, it
proves useful to define  the quantity $\hat{K}=K/k_{\rm{F}}=2
J_\Lambda /N$. It corresponds to the overall momentum in units of
$k_{\rm F}$ and measures the position, where the extra particle's
momentum is located in the Fermi--sea . For $\hat{K}=0$ the extra
particle's momentum is in the center, for $\hat{K}=\pm 1$ it is at
the upper or lower edge.

Since it
appears frequently troughout the following we introduced in
Eq.~\eqref{shift} the abbreviation
\begin{eqnarray}
v(k_{\rm F},c,K)=\left[\arctan\left(\frac{k_{\rm F}-\Lambda}{c}\right) +\arctan\left(\frac{k_{\rm F}+\Lambda}{c}\right) \right].
\end{eqnarray}
Note that the parameter $\Lambda$ in Eq.~(\ref{shift}) is not independent
but relates via the second Bethe-Ansatz equation (\ref{eq1.2.6})
to $J_{\Lambda}$, $k_{\rm F}$ and $c$
\begin{eqnarray}
J_\Lambda & = & \frac{1}{\pi}\int\limits_{-k_{\rm F}}^{k_{\rm F}}
dk \varrho(k)\arctan\left(\frac{k-\Lambda}{c}\right)\\
&=& \frac{N}{2\pi}\int\limits_{-1}^1 dx
\arctan\left(\frac{x-\Lambda/k_{\rm F}}{\hat{c}}\right) +
\mathcal{O}(N^0) \ . \nonumber
\end{eqnarray}
From this equation follows in particular that $\Lambda|_{c=0}= -2\pi J_\Lambda/L$.

Although the integral can be evaluated the resulting
transcendental equation can not be solved analytically  for
$\Lambda$. Therefore the energy shift can in general not be
expressed as function of the three independent quantities
$J_\Lambda$, $k_{\rm F}$ and $c$ and has to be calculated
numerically. For the ground state where $J_{\Lambda}=\Lambda=0$,
the situation is more favorable and the expression \eqref{shift}
simplifies to
\begin{eqnarray}
E-E_0&=&\frac{2}{\pi} \left[ c k_{\rm F}+ k^2_{\rm F}\arctan\left(\frac{c}{k_{\rm F}}\right) -c^2 \arctan\left(\frac{k_{\rm F}}{c}\right) \right]\label{eq1.2.1.2} .
\end{eqnarray}
The mean interaction energy
\begin{equation}
\langle \hat{V}\rangle \ = \ \langle \Psi_0| 4c \sum_{n=1}^N\delta(x_n-y)|\Psi_0 \rangle
\end{equation}
can be obtained using Pauli's trick (see Ref.~\cite{fet71}) as
 \begin{eqnarray}\label{va}
 \langle \hat{V}\rangle & = & c \frac{dE}{d c}\ = \  \frac{4c}{\pi}\left[k_{\rm F} - c \arctan\left(\frac{k_{\rm F}}{c}\right)\right] \ .
\end{eqnarray}
The interaction energy is given by the interaction strength times
the probability to find a sea-particle at the same position as the
extra particle. The latter is identical with the equal time
density--density correlation function $R(x,y)$ at $x = y=0$. Thus
we have, by definition,
\begin{equation}\label{vb}
\langle \hat{V}\rangle \ = \ 4c\, L \, R(0,0) \ .
\end{equation}
Equating Eqs.~(\ref{va}) and (\ref{vb}) yields
\begin{equation}
R(0,0) \ =\ \frac{N}{\rho}\left[ 1 - \hat{c} \arctan\left(\frac{1}{\hat{c}}\right)\right]\label{r0c0} .
\end{equation}
Thus the probability to find a sea particle at the position of the
extra particle vanishes as $c^{-2}$ for $c\to\infty$. Therefore
the interaction energy vanishes in the weak coupling limit as well
as in the strong coupling limit.

\begin{figure}[ht!]
    \centering
 \includegraphics[scale=0.9]{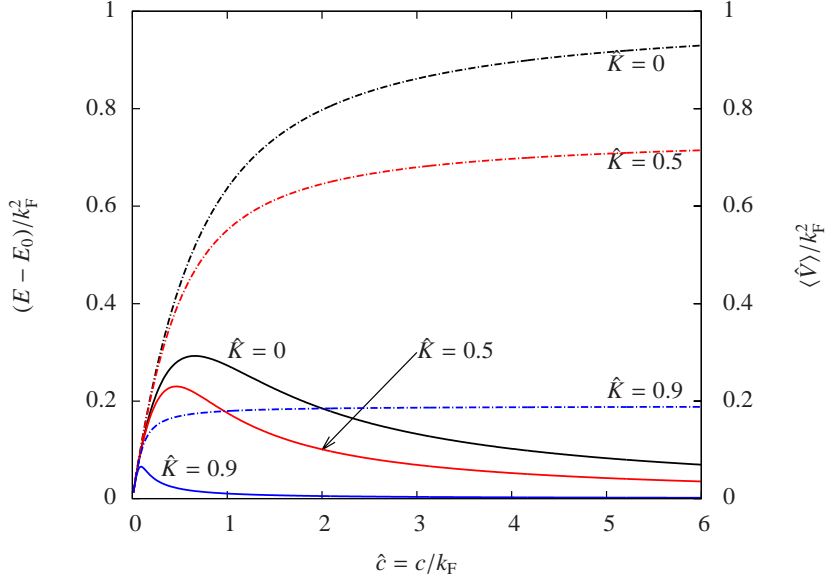}
    \caption{\label{fig:shift} Interaction energy (full lines) and  energy shift (dot-dash lines) as
    function of $ \hat{c}$. The values of
    $\hat{K}$ are $\hat{K}=0$ (black), $\hat{K}=0.5$ (online red)
 and $\hat{K}=0.9$ (online blue).  }
\end{figure}

In  Sec.~\ref{DD-corr}, where  the full density--density
correlation function is studied, we derive an expression for
$R(0,0)$ for $J_\Lambda \neq 0$, see Eq.~\eqref{eq2.9}.  This
allows us to calculate the interaction energy in this case as
well.  In Fig.~ \ref{fig:shift} the ground state energy shift and
interaction energy are plotted versus the interaction strength for
three different values of $\hat{K}$.
It is seen that the interaction energy has a maximum as predicted.
The ground state energy shift increases monotonously and saturates
at a value $E_{\rm max}=$ $k_{\rm F}^2-K^2$.  This means that
for the extra particles momentum located right at the edge of the
Fermi--sea the energy shift vanishes for arbitrary interaction
strength.

Using the wave function \eqref{eq1.1.2} the mean expectation value
of the extra particles energy can  be calculated. Up to
corrections of order $1/N$ we obtain
\begin{eqnarray}
\langle \hat{p}_y^2 \rangle &=&\Lambda^2 -c^2 + v^{-1}(k_{\rm F},
c, K) \left( 2k_{\rm F}c+ c\Lambda \ln \left( \frac{c^2
(k_{\rm F}-\Lambda)^2}{c^2 (k_{\rm F}+\Lambda)^2}\right)\right).
\end{eqnarray}
For the ground state this simplifies to
\begin{eqnarray}
\langle \hat{p}_y^2 \rangle  =k_{\rm F}^2 \hat{c}\left(\frac{1}{\arctan(\hat{c}^{-1})} -\hat{c}\right).\label{eq1.2.1.7}
\end{eqnarray}
From Eq.~\eqref{eq1.2.1.7} it is seen that the extra particles
energy increases monotonically from $\langle \hat{p}_y^2
\rangle=0$ to $\langle \hat{p}_y^2 \rangle= k_{\rm F}^2/3$ as
$\hat{c}$ varies from zero to $+\infty$.

\section{Equal Time Green's Function}\label{green}

The equal time single particle Green's function with respect to a
wave function $\Psi$ is defined as the $N$-fold integral
\begin{eqnarray}
G(y,y^\prime)&=& \int\limits_0^{L}
dx_1\cdots\int\limits_0^{L} dx_{N}
\Psi(x_1,\ldots,x_N,y)\Psi^*(x_1,\ldots,x_N,y^\prime) \label{eq1.1.5} .
\end{eqnarray}
Likewise it can be interpreted as the reduced density matrix of
the extra particle. We stick to the notation {\em equal time
Green's function} or simply \emph{Green's function}. Using the
representation \eqref{eq1.1.2}  for the wave function $\Psi$,
closed expressions can be obtained for $G(y,y^\prime)$. Expanding
the determinant in Eq.~\eqref{eq1.1.2}  with respect to the last
column and using the normalization condition $G(0,0)=1/L$ we
obtain
\begin{eqnarray}
G(y,y^\prime)&=&\left|C_N\right|^2 \sum_{n=1}^{N+1}\sum_{m=1}^{N+1}(-1)^{n+m} e^{\imath(k_ny -k_my')}\label{eq3.1.0}\\
&&\times\int\limits_0^L dx_1\ldots\int\limits_0^L dx_N \det\left[ A_j(x_l-y)e^{\imath k_j x_l}\right]_{\substack{j=1,\ldots,N+1\neq n\\ l=1,\ldots,N}}\nonumber\\
&&\hspace*{2cm}\times\det\left[A^*_j(x_l-y^\prime)e^{-\imath k_j
x_l}\right]_{\substack{j=1,\ldots,N+1\neq m\\
l=1,\ldots,N}}\nonumber.
\end{eqnarray}
Employing the general result
\begin{eqnarray}
&&\int dx_1\ldots \int dx_N \det\left[f_j(x_l)\right]_
{j,l=1,\ldots,N}\det\left[h_j(x_l)\right]_{j,l=1,\ldots,N}\label{eq2.51}\\
&&\qquad\qquad\qquad\qquad =\ N!\det\left[\int dx h_j(x)f_l(x)\right]_{j,l=1,\ldots,N}\nonumber,
\end{eqnarray}
one determinant can be replaced by its diagonal part and the
$x$--integrations can be performed. The resulting expression
acquires the form
\begin{eqnarray}
G(y,y^\prime)&=&\left|C_N\right|^2N! \sum_{n=1}^{N+1}\sum_{m=1}^{N+1}(-1)^{n+m}
e^{\imath(k_n +k_m)(y-y^\prime)/2}\nonumber\\
&&\hspace*{3.5cm}\times\det\left[K_{jl}(y,y^\prime)\right]_{\substack{j=1,\ldots,N+1 \neq
n\\l=1,\ldots,N+1 \neq m}}\label{eq3.1.2},
\end{eqnarray}
where the matrix entries are given by
\begin{eqnarray}
&&\hspace*{-0.4cm}K_{jl}(y,y^\prime)=e^{-\imath(k_j -k_l)(y+y^\prime)/2}\int\limits_0^L dx A_j(x-y)A^*_l(x-y^\prime)e^{\imath(k_j -k_l)x} \label{eq3.1.3} .
\end{eqnarray}
The evaluation of $K_{jl}(y^-)$ is straightforward but tedious. It
yields
\begin{eqnarray}
K_{jl}(y^-)&=& \left[L((k_j -\Lambda)^2+c^2) +2c\imath(k_j -\Lambda)y^- -2c^2y^-)\right]\delta_{jl}\label{eq3.1.4}\\
 &-&2c\left\{\cos \left(\frac{(k_j -k_l)y^-}{2}\right) \nonumber\right.\\
 &&\left.\qquad -\imath \frac{k_l + k_j - 2\Lambda+i2c\sgn(y^-)}{k_j -k_l}\sin\left(\frac{(k_j -k_l)y^-}{2}\right)\right\}\left(1-\delta_{ij}\right)\nonumber\
\end{eqnarray}
with  $y^{-}= (y - y^\prime)$. The expression \eqref{eq3.1.4}
makes explicit that $G(y,y^\prime)$ is a function of the
difference $y^-$ only, as expected by translation invariance.
After some further matrix algebra the  r.~h.~s.~of
Eq.~\eqref{eq3.1.2} can be can be expressed in terms of a single
$N\times N$ determinant
\begin{eqnarray}
G(y,y^\prime)  &=& \frac{e^{\imath Ky^-}}{L}G_{\rm{I}}(y^-)\label{greendecom1},
\end{eqnarray}
where we have introduced the interaction part $G_{\rm{I}}(y^-)$ of
the Green's function
\begin{eqnarray}
G_{\rm{I}}(y^-)= \left|C_N\right|^2N! L \det \begin{bmatrix}g_{nm}- g_{nm+1}-g_{n+1m} +g_{n+1
m+1}\end{bmatrix}_{n,m=1,\ldots,N}\ \label{greendecom}
\end{eqnarray}
and the quantities
\begin{eqnarray}
 g_{nm}=e^{-\imath(k_n +k_m)y^-/2}K_{nm}(y^-).\label{eq3.1.6}
\end{eqnarray}
According to Eq.~\eqref{greendecom1} we have $G_{\rm{I}}(0)=1$.
Separating the overall momentum  and $G_{\rm{I}}(y^-)$ as in Eq.~\eqref{greendecom1} is useful when exited states with $K\neq 0$ are studied. The representation of the Green's
function as a determinant is most convenient for a further numerical analysis.

\begin{figure}[ht!]
    \centering
  \includegraphics[scale=0.9]{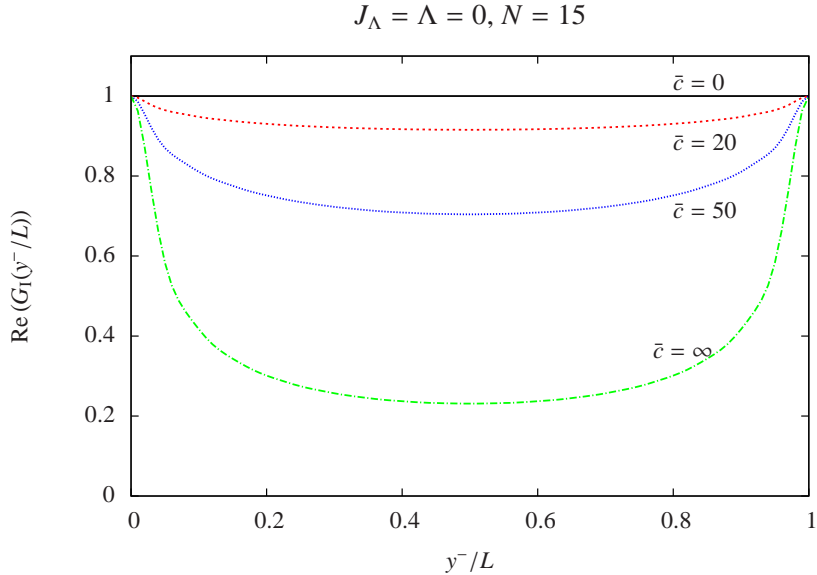}
    \caption{Real part of $G_{\rm{I}}(y^-)$   as function $y^-/L$
    for the ground state i.e.  $J_{\Lambda}=0$ and $N=15$. The values of
       $\bar{c}$  are $\bar{c}=0$ (full line, black),
 $\bar{c}=20$ (dashed line, online red),
       $\bar{c}=50$ (dotted line, online blue) and $\bar{c}=\infty$ 
(dot-dashed line, online green).   }
    \label{Green1}
\end{figure}

\begin{figure}[ht!]
    \centering
  \includegraphics[scale=0.9]{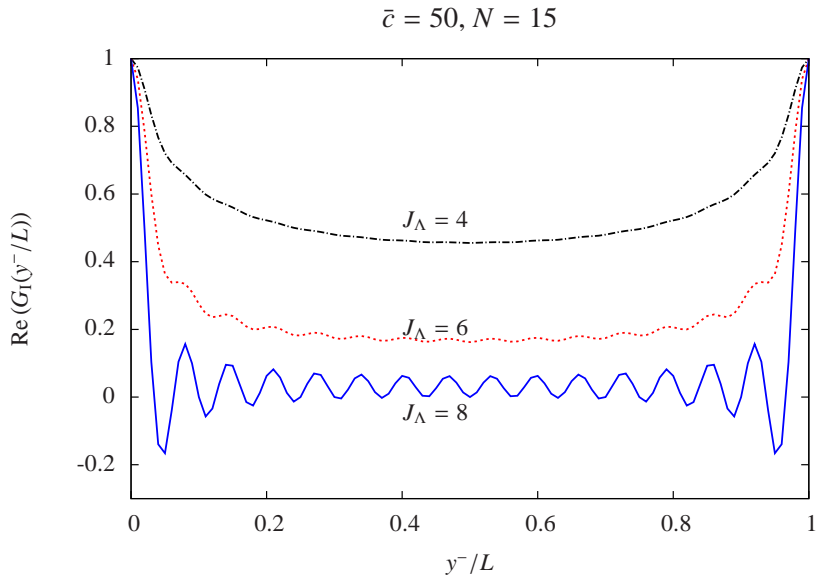}
    \caption{Real part of $G_{\rm{I}}(y^-)$   as function $y^-/L$ for $\bar{c}=50$
    and $N=15$. The values of   $J_{\Lambda}$ are  
 $J_{\Lambda}=4$ (dot-dashed line, black),
      $J_{\Lambda}=6$ (dotted line, online red), and $J_{\Lambda}=8$ 
(full line, online blue). }
    \label{Green2}
\end{figure}

In Figs.~\ref{Green1} and \ref{Green2}  the real part of
$G_{\rm{I}}(y^-)$ is plotted for a particle number of $N=15$. The
quantum numbers are chosen according to Eq.~\eqref{ms} and  $\bar{c}=cL$.
Figure~\ref{Green1} shows the transition of the interaction part of the Green's
function for the ground state  as $\bar{c}$ varies from
$\bar{c}=0$ to $\bar{c}=\infty$, and Fig. \ref{Green2} shows
 $G_{\rm{I}}(y^-)$ for exited states with
$J_{\Lambda} \neq 0$ and fixed $\bar{c}=50$.
Whereas in the ground state the curve is smooth and decays monotonously up to $y^-/L=0.5$, for  excited
states wiggles develop and the function seems to become oscillatory. However, only for the highest
value $J_\Lambda=8$, i.~e. for the particle just outside the Fermi-sea, the curve has nodes.

In the following, we analyze the representation \eqref{greendecom}  of
$G_{\rm{I}}(y^-)$ further in  the hardcore limit. We show that the features, shown in
Figs.~\ref{Green1} and \ref{Green2}  and described above persist in the thermodynamic limit.

\subsection{Hardcore Limit }\label{sec3.1}

For $c\rightarrow \infty$ the wave function~\eqref{eq1.1.2}
becomes
\begin{eqnarray}
 \Psi(x_1,\ldots,x_N,y)\propto
\prod_{j=1}^N \Big(-\imath \lambda + \sgn(x_j - y)
\Big)\det\left[e^{\imath k_j x_l}\Big|e^{\imath
k_jy}\right]_{\genfrac{}{}{0pt}{2}{j=1,\ldots,N+1}{ l=1,\ldots,N}},\nonumber\\
\label{eq1.1.3}
\end{eqnarray}
where we recall $\lambda= \Lambda/c$. For $\lambda=0$ the
expression resembles the wave function of $N+1$ hardcore Bosons
\cite{lie63}. The only difference consists in the product of
sign--functions: The wave function for hardcore Bosons is a Slater
determinant multiplied not with a single product of sign-functions
but with a double product $\prod_{i<j}\sgn(x_i-x_j)$. For the
equal time Green's function this difference is irrelevant and it
is identical with those of hardcore Bosons. A large body of
literature has been devoted to the studies of the latter
\cite{len64,len72,vai79,vai79a,jim80,for02,for03,for03a,for06,pap03}.

On the other hand for $|\lambda| \rightarrow \infty$ the wave
function and therefore also the Green's function coincides with
that of free Fermions. This is already reflected in the
Bethe-Ansatz equations \eqref{eq1.2.9} and \eqref{eq1.2.10}. The
quasimomenta  are integer multiples of $2\pi/L$ with an offset
$\arctan(\lambda)/\pi$ which ranges from zero to
$\pm{\scriptstyle\frac{1}{2}}$ for $|\lambda|\rightarrow \infty$.

Hence varying $\lambda$ from $\lambda=0$ to $\lambda=\infty$ we
expect the extra particle to undergo a transition from a hardcore
Boson to a free Fermion.
 This transition should be reflected in the equal time Green's function.
As there are two fundamental formulations of the Green's function
for impenetrable Bosons we will go for two different ways to study
it. The first approach is based on the theory of  T{\"o}plitz
determinants \cite{len64,len72,wid73,for02} and will be reviewed in the appendix \ref{appA}. 

The second approach is the description of the Green's function via solutions of a Painlev{\'e} equation \cite{jim80,for03}. In the following we are going to show that the
above mentioned transition is for zero temperature described by a
change in the initial condition of the solution of one and the
same Painlev{\'e} equation.

We plug Eq.~\eqref{eq1.1.3} into the definition   \eqref{eq1.1.5}
of the equal time Green's function. This yields
\begin{eqnarray}
G(y,y^\prime)&=&\left|C_N\right|^2\int d[x] \prod_{l=1}^{N} (\imath \lambda +\sgn(x_l -y) )(-\imath \lambda +\sgn(x_l-y^\prime))\nonumber\\
&&\hspace*{-0.5cm}\times\det\left[e^{\imath k_j x_l}\Big|e^{\imath
k_jy}\right]_{\substack{j=1,\ldots,N+1\\ l=1,\ldots,N}}
\det\left[e^{-\imath k_j x_l}\Big|e^{-\imath
k_jy^\prime}\right]_{\substack{j=1,\ldots,N+1\\ l=1,\ldots,N}}\label{eq3.2.12} \ .
\end{eqnarray}
The crucial point is that for zero temperature, or -- more
precisely -- if the Fermi--sea is in the ground state, the
integral can be interpreted  as an average over  the unitary
group. We rewrite the last line of the equation above according to
\begin{eqnarray}
\lefteqn{\det\left[e^{\imath k_j x_l}\Big| e^{\imath
k_jy}\right]_{\substack{j=1,\ldots,N+1\\ l=1,\ldots,N}}\det\left[e^{-\imath k_j x_l}\Big| e^{-\imath
k_jy^\prime}\right]_{\substack{j=1,\ldots,N+1\\ l=1,\ldots,N}} }\label{eq3.2.18}\\
&&\qquad\propto \exp\left( \frac{2\imath}{L}\arctan(\lambda)y^-\right)
\prod_{1\leq j<l\leq N}\sin^2\left(\frac{\pi(x_j-x_l)}{L}\right)\nonumber\\
&&\qquad \qquad\qquad\quad\qquad\prod_{l=1}^N\sin\left(\frac{\pi x_l}{L} - \frac{\pi y}{L} \right)\sin\left(\frac{\pi x_l}{L} - \frac{\pi y^\prime}{L} \right) . \nonumber
\end{eqnarray}
In the present case the first product can  be identified with the measure of the unitary
group $U(N)$ \cite{meh04}
\begin{equation}
d\mu(U) \ \propto \prod_{1\leq j<l\leq N}\sin^2\left(\frac{\pi(x_j-x_l)}{L}\right) \prod_{j=1}^N\frac{2\pi d x_j}{L} \   .
\end{equation}
The Green's function can again  be written like in  Eq.~\eqref{greendecom1}, where the part $G_{\rm{I}}$, which
includes the effects of interaction, is identified with the
average of the function
\begin{equation}
\label{measurefunc}
 \prod_{l=1}^N\left(1-\xi \chi^{(l)}_{[y^\prime,y)}\right)2\sin\left(\pi x_l\right)\left(-e^{-\frac{\pi \imath}{L} \left(x_l- y^-\right)}\right)\left(1- e^{\frac{2\pi \imath}{L} \left( x_l- y^-\right)}\right)
\end{equation}
with respect to the measure $d\mu(U)$.  Here we  have introduced the indicator function
\begin{eqnarray}
\chi^{(l)}_{[y',y)}=\begin{cases}
1 \quad \textrm{for}\quad x_l\in[y^\prime,y)\\
0 \quad \textrm{else}
\end{cases}
\label{eq3.2.16}
\end{eqnarray}
and the parameter
\begin{eqnarray}
\xi=\frac{2}{1 +\imath \lambda} \label{eq3.2.17}.
\end{eqnarray}
Averages over the unitary group of functions of the type
\eqref{measurefunc} were studied by Forrester, Frankel, Garoni and Witte
\cite{for01,for02,for03} and they have been related to solutions
of the Painlev{\'e} {\RM 6} non--linear differential equation.
Defining
\begin{equation}
u \ =\ \exp\left(-\imath\frac{2\pi  y^-}{L}\right),
\end{equation}
we follow Proposition 3 and Corollary 1 of
Ref.~\cite{for03} in order to deduce the relation
\begin{eqnarray}
u(u-1)\frac{d}{du}\ln G_{\rm I}(y^-)\Big|_{u=e^{-\imath 2\pi
y^-/L}}= \sigma_{N+1}(u) \label{eq3.2.21}
\end{eqnarray}
for the interaction part of the Green's function. The function
$\sigma_{N}(u)$ is a solution of  the particular $P_{\rm \RM{6}}$
$\sigma$-form due to Okamato \cite{oka86}
\begin{eqnarray}
- u^2(u-1)^2(\sigma_N^{\prime\prime})^2 &=& \left(\sigma_N - (u-1) \sigma^\prime_N +1\right)\label{eq3.2.22} \\
&& \times\left[4 \sigma^\prime_N(\sigma_N -u\,  \sigma^\prime_N)- (N^2 -1)(\sigma_N-(u-1)\sigma^\prime_N)\right]\nonumber.
\end{eqnarray}
The  initial conditions are fixed by expanding $G_{\rm I}(y^-)$
for small distances like in Ref.~\cite{for03}. We find that in
terms of the variable $u$ the solution of Eq.~\eqref{eq3.2.22},
quested for, has the small distance expansion
\begin{eqnarray}
\sigma_{N}(u) &=& \frac{N^2 -1}{12}(u-1)^2 \label{eq3.2.26}\\
 &&\qquad \qquad+ \frac{\left(N^2 -1\right) \left( \imath N/(1 +\imath \lambda) - \pi\right)}{
24 \pi} (u-1)^3 +\ldots \nonumber\quad.
\end{eqnarray}
In the thermodynamic limit the expressions simplify further.
First, Painlev{\'e} {\RM 6} converts  in the limit $N\to\infty$ to
the Jimbo--Miwa--Okamoto form \cite{jim80,oka87} of the
Painlev{\'e} {\RM 5} differential equation
\begin{eqnarray}
\frac{x^2}{4}(\sigma_{\RM 5}^{\prime\prime})^2 &=&
\left(\sigma_{\RM 5}- x \sigma_{\RM 5}^\prime +1\right)\left(
(\sigma_{\RM 5}^\prime)^2 - \sigma_{\RM 5} + x
 \sigma_{\RM 5}^\prime \right)\label{pain5}\ ,
\end{eqnarray}
where $x=k_{\rm F} y^-$. Second, the initial condition for
$\sigma_{\RM 5}(x)$ becomes
\begin{equation}
\label{init}
\sigma_{\RM 5}(x) \ \sim \ -\frac{x^2}{3}+\frac{1}{1+\imath \lambda}
\frac{x^3}{3\pi}+\ldots  \ ,\qquad x\to 0  \ .
\end{equation}
Thus the Green's function reads in the thermodynamic limit
\begin{eqnarray}
\label{greentherm}
G(x) & = & \frac{e^{\imath \hat{K} x}}{L} G_{\rm{I}}(x) \, \nonumber\\
G_{\rm{I}}(x) & = & \exp\left(\imath x\hat{K}+\int_0^x dx^\prime \frac{\sigma_{\RM
5}(x^\prime)}{x^\prime}\right) \ .
\end{eqnarray}
The long distance behavior of $G_{\rm{I}}(x)$  can be related to
the short distance behavior (\ref{init}), i.~e. to $\lambda$ with
a connection formula  for Painlev\'e equations. The case of
Painlev{\'e} \RM{5} has been solved  by McCoy and Tang
\cite{mcc86a,mcc86b,mcc86c}.  We employ the results
obtained in Ref.~\cite{mcc86b} for hardcore Bosons:

Let the solution  of Painlev\'e {\RM 5} in the Jimbo-Miwa-Okamoto
sigma form as given in Eq.~(\ref{pain5}) be regular at $x=0$ with
the expansion
\begin{equation}
\label{inita}
\sigma_{\RM 5}(x) \ \sim \ -\frac{x^2}{3} + \frac{\xi} {6\pi} \, x^3 +\ldots  \ ,\qquad x\to 0  \ .
\end{equation}
Then for general $\xi\in{\mathbb C}$ the asymptotic expansion for
$x\to\infty$ is different for  $\xi \in (1,\infty)$ (case I) and
for $\xi \in {\mathbb C}\setminus (1,\infty)$ (case II). In the
first case it its given by
\begin{eqnarray}
\sigma_{\RM 5}(x) & = & 2 k_{\rm I} x + 2k_{\rm I}^2-\frac{1}{2}+\label{case1a}\\
  &&  \frac{1}{4 x}\left[\left(4k_{\rm I}^2-1\right)\sin 2 s_{\rm I}(x) +
  2 k_{\rm I} \cos 2s_{\rm I}(x) - 2 k_{\rm I} \left(4k_{\rm I}^2+1\right)\right] \nonumber\\
 && - \frac{1}{x^2}\left[ k_{\rm I} \left(4 k_{\rm I}^2+1\right)^2
 \sin 2 s_{\rm I}(x) - M_{\rm I}\right] + {\cal O}(x^{-3})\ ,\nonumber\\
 s_{\rm I} (x) &=& x + \phi_{\rm I} + 2k_{\rm I} \ln\left(x\right) \ ,
 \label{case1}
\end{eqnarray}
where the parameters $k_{\rm I}, \phi_{\rm I}$  depend on $\xi$ as
\begin{eqnarray}
k_{\rm I} & = & \frac{1}{2\pi}\ln\left(\xi-1\right) \ ,\\
e^{2\imath \phi_{\rm I}} & = & - 2^{4\imath k_{\rm I}} \frac{\Gamma^2(-\imath k_{\rm I}+1/2)}
{\Gamma^2(\imath k_{\rm I}+1/2)}\nonumber
\end{eqnarray}
and $M_{\rm I}$ is a constant to be determined.
In case II the asymptotic expansion is
\begin{eqnarray}
\sigma_{\RM 5}(x) & = & x\left(\cot s_{\rm II}(x) + 2 k_{\rm II}\right) \label{case2a} \\
&&+ \frac{1}{2}\left(4 k_{\rm II}^2-2\right)
+ \frac{3 k_{\rm II}^2+1}{\left(\sin s_{\rm II}(x)\right)^2} +  2 k_{\rm II} \cot s_{\rm II}(x) \nonumber\\
 && + \frac{1}{x}
\left[\frac{(1+3k^2_{\rm II})^2\cot s_{\rm II}(x)}
{(\sin s_{\rm II} (x))^2} + (1+3k^2_{\rm II})
\cot s_{\rm II} (x) + \right. \nonumber\\
&&\qquad\qquad\left. \frac{L_{\rm II}}
{(\sin s_{\rm II} (x))^2}  +
M_{\rm II}\right] + {\cal O}(x^{-2}) \ ,\nonumber\\
s_{\rm II} (x) &=& x + \phi_{\rm II} + 2 k_{\rm II}
\ln\left(x\right) \ , \label{case2b}
\end{eqnarray}
where $L_{\rm II}$, $M_{\rm II}$ are unknown constants. The
parameters are now\footnote{There is a typo in Eq.~(1.29c) of
Ref.~\cite{mcc86b}. The square bracket has to read
$[\theta+\frac{A^2}{4}-n^2-1]$.}
\begin{eqnarray}
k_{\rm II} &=& \frac{1}{2\pi}\ln\left(1-\xi \right) \ ,\\
e^{2\imath \phi_{\rm II}} & = & 2^{4\imath k_{\rm II}}
\frac{\Gamma^2(-\imath k_{\rm II})}{\Gamma^2(\imath k_{\rm
II})} \ .\nonumber
\end{eqnarray}
Only for $\lambda= 0$,  $\xi \in (1,\infty)$ and case I applies.
As mentioned above for $\lambda = 0$ the extra particle's Green's
function becomes identical with the single particle Green's
function of a system of identical hardcore Bosons. For that case
Vaidya and Tracy \cite{vai79,vai79a} derived an asymptotic
expansion, whose first terms read
\begin{equation}\label{asymptotics}
G_{\rm{I}}(x) \ = \
\frac{G_\infty}{\sqrt{x}}\left\{1+\frac{1}{8x^2}\left[\cos(2x)-
\frac{1}{4}\right]+{\cal O}(x^{-3})\right\} \ .
\end{equation}
This result can not completely be derived  from Eqs.~(\ref{case1a}) and (\ref{case1}). The constant $G_\infty\approx
\pi\sqrt{e} 2^{-1/3}A^{-6}$, where $A\approx1.2842$ is Glaisher's
constant, cannot be obtained from  the asymptotic expansion of
$\sigma_{\RM 5}$. Neither can the constant term in the square bracket. Rather Eq.~(\ref{asymptotics}) fixes the
constant $M_{\rm I}=2^{-4}$.

For $|\lambda|> 0$ case II applies.  We write $k_{\rm II}= \imath \alpha$, where
\begin{equation}
\alpha = \frac{\sgn(\lambda)}{2} - \frac{1}{\pi}\arctan(\lambda)
\end{equation}
is a number between $-1/2$ and $1/2$.
The cases $\alpha = 0$ and $\alpha \neq 0$ have to be treated separately.
 By expanding Eq.~(\ref{case2b}) for $\alpha =0$ up to leading
 order we obtain $\sigma_{\RM 5}\sim$ $x \cot(x)-1$. 
This together with $\hat{K}=0$ yields 
\begin{equation}
\label{free}
G_{\rm I}(x) \ =\ e^{\imath x}\frac{\sin(x)}{x} \ .
\end{equation}
This is the well known Green's function of a system of free Fermions. Likewise,
it could have been derived from Eq.~(\ref{eq1.1.3})  in the limit $\lambda\to\infty$.
It is easily checked that $x \cot(x)-1$ is  an exact solution to the Painlev{\'e} V
differential equation (\ref{pain5}) to the boundary
condition $\lim_{x\to 0}\sigma_{\RM 5} = -x^2/3$.

For $\alpha\neq 0$ the asymptotic expansion yields
\begin{eqnarray}
\label{asympalneq0}
G_{\rm I}(x) & = & G^{(\lambda)}_\infty x^{-(\hat{K}^2+1)/2}\left(1-\frac{\Gamma^2(\alpha)}{\Gamma^2(-\alpha)}\frac{e^{2\imath x}}{(2x)^{4|\alpha|}} -
\frac{M_{\rm II}-\imath (1-3\alpha^2)}{x}\right.\nonumber\\
& &\qquad \left.+{\cal O}(x^{-4|\alpha|-{\rm min}(4 |\alpha|,1)})\right) \ .
\end{eqnarray}
The exponent of the leading order term can be written as a
function of the extra particle's momentum $\hat{K}$. It
interpolates between $-1/2$ for the extra particles momentum in
the core of the Fermi--sea and $-1$ for the extra particle's
momentum right at the edge of the Fermi--sea. As $\alpha$
approaches zero the approximation in Eq.~(\ref{asympalneq0})
becomes poorer and poorer. Finally for $\alpha=0$ infinitely many
terms contribute to the same order and sum up to the simple result
(\ref{free}). Within the present approach the constants 
$G^{(\lambda)}_\infty$ and $M_{\rm II}$
of the asymptotic expansion can not be determined. The first subleading term in the
expansion is of order $\min(4|\alpha|, 1)$.

\begin{figure}[ht!]
    \centering
\includegraphics[scale=1]{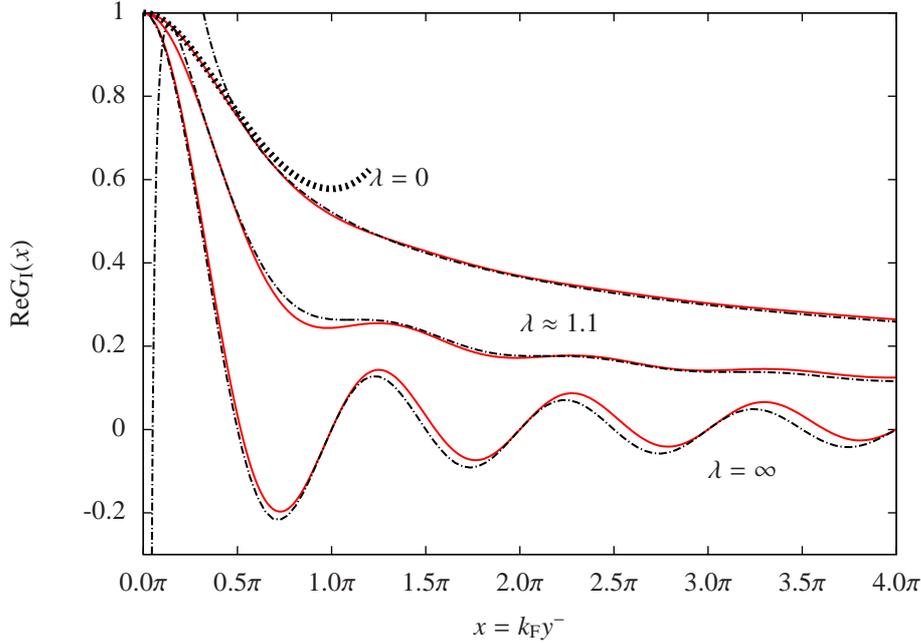}
    \caption{Real part of $G_{\rm{I}}(y^-)$ for $N=29$.
    The values of $\lambda$ are $\lambda=0$,
    $\lambda\approx 1.1$ and $\lambda=\infty$.
    The thin dot-dashed lines correspond
    to the large $x$ expansion (black). 
    The thick dotted line (black) corresponds to the small $x$
    expansion for $\lambda=0$.
    The full lines (online red) are obtained from Eq.~\eqref{eq3.2.7-1}. }
    \label{fig.GreenTHL2}
\end{figure}

In Fig. \ref{fig.GreenTHL2}  $G_{\rm{I}}(y^-)$ is plotted for
different values of  $\lambda$. The curves are generated from the
representation of $G_{\rm{I}}(y^-)$ as a T\"oplitz determinant
(see Eq.~\eqref{eq3.2.7-1} in the appendix~\ref{appA}). The number of
sea particles is $N=29$. The expansion for large distances in the
thermodynamic limit as given in Eqs.~\eqref{init},\eqref{asymptotics} and (\ref{asympalneq0}) are compared with the
finite $N$ result. The constants  $ G^{(1.1)}_\infty \approx 0.6$
and $M_{\rm II}\approx 0$ were numerically approximated.
In particular for the non---oscillatory curve for $\lambda=0$ 
the asymptotic expression works remarkably well almost everywhere.

\subsection{Finite Interaction Strength}\label{sec3.2}

Up to now we were unable to  evaluate the Green's function in the
thermodynamic limit for finite interaction strength. Nevertheless
the representation \eqref{greendecom} of $G_{\rm I}(y^-)$ allows us to
deduce the small distances behavior.
We expand $G_{\rm{I}}(y^-)$  for small $y^-$. As starting point we
take the quantities $g_{nm}$ defined in Eq.~\eqref{eq3.1.6}.
Expanding them up to the quadratic term yields
\begin{eqnarray}
g_{nm}&=& (b_n \delta_{nm} -2c) +
\left( \imath(- k_nb_n\delta_{nm} +2c(k_n+k_m-\Lambda))
 -2c^2\right)y^-\label{eq4.1}\\
&&\ + \frac{1}{4}\left[-2k_n^2 b_n
\delta_{nm}+4c(k_n^2 +k_m^2 +k_n k_m)\right.\nonumber\\
&&\ \left.-4c(k_n+k_m)(\Lambda -\imath c)\right]\left(y^-\right)^2 + \mathcal{O}\left((y^-)^3\right) \ ,\nonumber
\end{eqnarray}
where we have introduced the notation
\begin{eqnarray}
b_n=L[(k_n-\Lambda)^2 +c^2] +2c \ . \label{eq4.2}
\end{eqnarray}
For the linear combination of the quantities $g_{nm}$ as it
appears in the Eq.~\eqref{greendecom} we obtain
\begin{eqnarray}
&&\hspace*{2cm}g_{nm} -g_{n+1m}-g_{nm+1}+g_{n+1m+1}\label{eq4.3}\\
&=&\left[(b_n +b_{n+1})\delta_{nm}- b_n\delta_{nm+1}
-b_{n+1} \delta_{n+1m}\right]\nonumber\\
&-&\imath \left[(b_n k_n +b_{n+1}k_{n+1})\delta_{nm}-
b_n k_n\delta_{nm+1} -b_{n+1}k_{n+1} \delta_{n+1m}\right]y^-\nonumber\\
&-&\frac{1}{2}\Big(\left[(b_n k_n^2
+b_{n+1}k^2_{n+1})\delta_{nm}- b_n k^2_n\delta_{nm+1}
-b_{n+1}k^2_{n+1} \delta_{n+1m}\right]\nonumber\\
&&-2c(k_n-k_{n+1})(k_m-k_{m+1})\Big)
(y^-)^2+\mathcal{O}((y^-)^3) \ .\nonumber
\end{eqnarray}
We plug the expansion \eqref{eq4.3} into Eq.~\eqref{greendecom}
and write $G_{\rm{I}}(y^-)$ formally in a power series  up to the quadratic term
\begin{eqnarray}
G_{\rm{I}}(y^-)&=& 1+\imath G_{\rm{I}}^{(1)}y^--G_{\rm{I}}^{(2)}(y^-)^2+
\mathcal{O}\left((y^-)^3\right)\label{eq4.4} \ .
\end{eqnarray}
Our numerical results (see Figs.~\ref{Green1} and \ref{Green2}) 
together with the results obtained in the hardcore limit indicate 
that $G_{\rm{I}}(y^-)$  is nodeless for $K<k_{\rm F}$. Therefore the
short distances decay of the Greens function is dominated 
by the length $[G_{\rm{I}}^{(2)}]^{-1/2}$
to which we refer as correlation length.
In Eq.~\eqref{eq4.4} we used that $G_{\rm{I}}(0)=1$ by definition. 
Comparing with Eq.~\eqref{greendecom} we have
 \begin{eqnarray}
(N!L \left|C_N\right|^2)^{-1}= \det\left[ (b_n +b_{n+1})\delta_{nm}- b_n\delta_{nm+1} -
b_{n+1} \delta_{n+1m}\right]_{n,m=1,\ldots,N}.\nonumber\\\label{eq4.5}
\end{eqnarray}
The determinant on the r.~h.~s. can be evaluated
yielding for the normalization constant 
\begin{eqnarray}
 \left|C_N\right|^{-2}=LN!\left(\prod_{j=1}^{N+1} b_j \right)
\sum_{j=1}^{N+1}\frac{1}{b_j}.\label{eq4.7}
\end{eqnarray}
The evaluation of
the first and second order coefficient is somewhat lengthy but
straightforward. We obtain
\begin{eqnarray}
G_{\rm{I}}^{(1)}&=&\left(\sum\limits_{j=1}^{N+1}
\frac{1}{b_j}\right)^{-1}\left(\sum_{j=1}^{N+1}
\frac{K-k_j}{b_j}\right), \label{eq4.8}\\
G_{\rm{I}}^{(2)}&=&\frac{1}{2}\left(\sum\limits_{j=1}^{N+1}
\frac{1}{b_j}\right)^{-1}\left(\sum_{j=1}^{N+1}
\frac{(K-k_j)^2}{b_j}- c\sum_{j=1}^{N+1}
\sum_{l=1}^{N+1}\frac{(k_j -k_l)^2}{b_j b_l}\right)\label{eq4.9} .
\end{eqnarray}
By definition the antisymmetric part of $G(y^-)$ and
$G_{\rm{I}}(y^-)$ is purely imaginary while the symmetric part is
always real. This is reflected in the expansion \eqref{eq4.4} and in the expansion coefficients
\eqref{eq4.8} and \eqref{eq4.9}.
 In the thermodynamic limit the coefficients
\eqref{eq4.8} and \eqref{eq4.9} translate to
\begin{eqnarray}
G_{\rm{I}}^{(1)}&=&\frac{  c}{ v(k_{\rm F},c,K)}
\int\limits_{-k_{\rm{F}}}^{+k_{\rm{F}}}dk
\frac{(K-k)}{(k-\Lambda)^2 +c^2} \ ,\label{eq4.8-1}\\
G_{\rm{I}}^{(2)}&=& \frac{c }{2v(k_{\rm F},c,K)}
\left(\int\limits_{-k_{\rm{F}}}^{+k_{\rm{F}}}dk
\frac{(K-k)^2}{(k-\Lambda)^2 +c^2}\label{eq4.9-1}\right.\\
&&-\left.\frac{c}{2\pi}\int\limits_{-k_{\rm{F}}}^
{+k_{\rm{F}}}dk\int\limits_{-k_{\rm{F}}}^{+k_{\rm{F}}}dk'
 \frac{(k-k')^2}{[(k-\Lambda)^2 +c^2][(k'-\Lambda)^2 +c^2]}\right).\nonumber
\end{eqnarray}
The integrals  are elementary. However, the resulting
expressions give little insight.  Thus, we do not state them here.
Equating  $G_{\rm{I}}^{(1)}$ and $G_{\rm{I}}^{(2)}$ in the hardcore limit yields
\begin{eqnarray}
\lim_{c\rightarrow\infty} G_{\rm{I}}^{(1)} =  k_{\rm F} \hat{K} \ , 
\qquad \lim_{c\rightarrow\infty} G_{\rm{I}}^{(2)}=\frac{k^2_{\rm F}}{6}(1 + 3\hat{K}^2) 
\  .\label{eq4.15}
\end{eqnarray}
Note that Eq.~\eqref{eq4.15}
is in perfect agreement with the results of the Painlev{\'e}
transcendental evaluation of $G_{\rm{I}}(y^-)$ in the hardcore limit (compare
Eqs.~\eqref{init} and\eqref{greentherm}). On the other hand we
obtain in the limit of vanishing interaction strength
\begin{eqnarray}
\lim_{c \rightarrow 0} G_{\rm{I}}^{(1)} =  \lim_{c \rightarrow 0} G_{\rm{I}}^{(2)}=0 \label{eq4.12a}
\end{eqnarray}
where $\Lambda|_{c=0}=K$ has been used. According to the equation above 
we have $G_{\rm{I}}(y^-)=1$ for $c=0$.  This is reasonable since then the distinguishable particle does
not feel the presence of the Fermi-sea and is described by a
perfect coherent state.  Great simplification arises for the
ground state where $K=J_{\Lambda}=\Lambda=0$. As obvious from
Eq.~\eqref{eq4.8-1} the linear term then vanishes identically and
the correlation length acquires the simple form
\begin{eqnarray}
&&\hspace*{-0.5cm}[G_{\rm{I}}^{(2)}]^{-1/2}= \frac{1}{\sqrt{2}k_{\rm F}}\left[\frac{\hat{c}}{2\pi}\left(\frac{\pi}{\arctan(\hat{c}^{-1})}-2\right)\left(1- \hat{c}\arctan\left(\hat{c}^{-1}\right)\right)\right]^{-1/2}. \label{eq4.12}
\end{eqnarray}
In general  the correlation length  decays monotonically between the values given
 in Eq.~\eqref{eq4.12a} and \eqref{eq4.15} as $c$ varies from zero to $\infty$.
Comparing $[G_{\rm{I}}^{(2)}]^{-1/2}$ for the ground
state and exited states of the extra particle, it is found that it decreases faster with increasing $K$.

\begin{figure}[ht!]
    \centering
  \includegraphics[width=\textwidth]{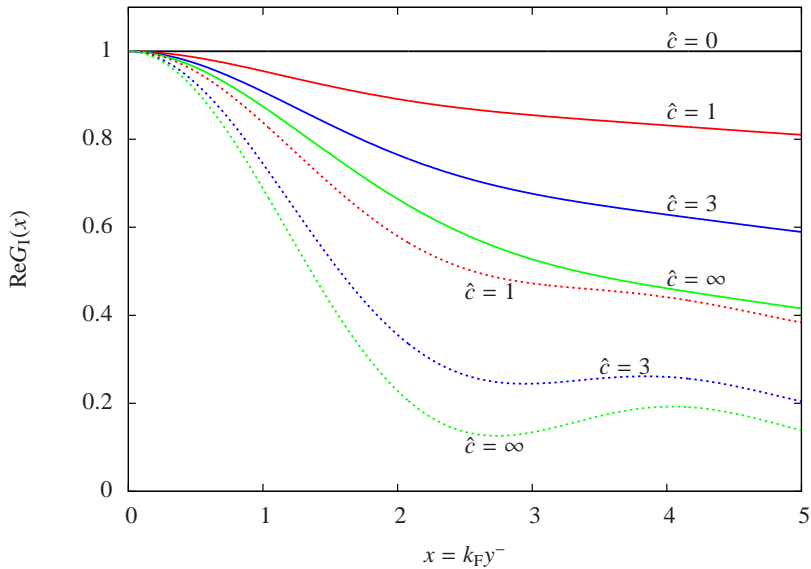}
    \caption{Real part of $G_{\rm{I}}(x)$  as function of
    $x=k_Fy^-$ for the ground state (solid lines) and for 
   a state with finite $\hat{K}\approx -0.6 $ (dotted lines). 
   The interaction strength $\hat{c}=c/k_{\rm F}$ has the values $\hat{c}=0$ 
  (black), $\hat{c}=1$ (online red), $\hat{c}=3$
    (online blue) and $\hat{c}=\infty$ (online green).
   The plots are generated for $N=29$ sea particles.
    \label{fig.GreenTHL1}}
\end{figure}

In Fig.~\ref{fig.GreenTHL1}  we show the
interaction part of the Green's function  for $J_{\Lambda}=
\hat{K}= 0$ and for an excited state with $\hat{K}= -0.6$ for different
values of the interaction strength. The particle number is $N=29$.
The decay is considerably faster for the excited state.

\section{Density-Density Correlation}\label{DD-corr}

The equal time density-density correlation function for a particle
of the Fermi-sea and the distinguishable particle has been
calculated in Ref.~\cite{mcg65}. However, there the results are
restricted to the case $J_{\Lambda}=\Lambda=0$. Here we give a
brief derivation of the corresponding quantity where the
distinguishable particle may occupy an exited state inside the
Fermi-sea.

\begin{figure}[ht!]
    \centering
  \includegraphics[scale=0.9]{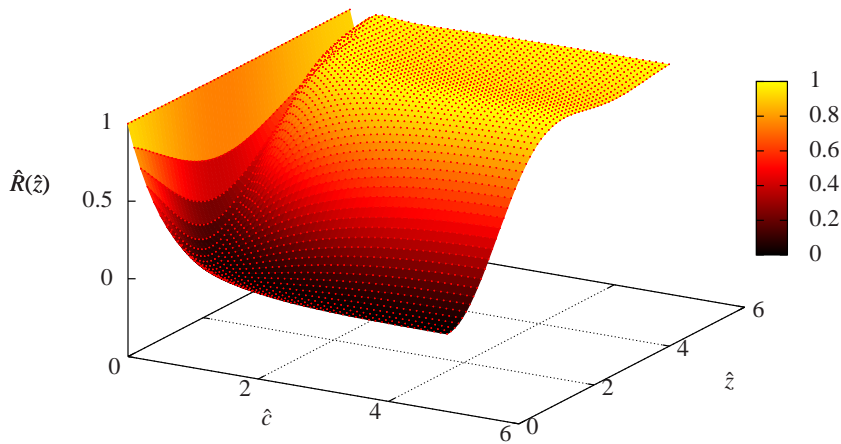}
\label{DD-corr3D}
    \caption{Density--density correlation $\hat{R}(\hat{z})$ as function of
    interaction strength $\hat{c}$ and distance
     $\hat{z}=z k_{\rm F}$ for the exited state with $\hat{K}=0.6$. Color 
online. }
\end{figure}

The density-density correlation function is defined by the
$(N-1)$-fold integral
\begin{eqnarray}
R(x,y)& = & N \int\limits_{0}^L dx_1 \cdots  \int\limits_{0}^L dx_{N-1}
 \left|\Psi(x_1,\ldots,x_{N-1},x,y)\right|^2.\label{eq2.1}
\end{eqnarray}
Using the wave function \eqref{eq1.1.2} the above expression is
after some algebra rewritten as
\begin{eqnarray}
R(z)&\propto&\sum_{ n=1}^{N+1} \frac{N}{\Big|A_n(1)\Big|^2}-\sum_{
n\neq m}^{N+1} \frac{e^{\imath(k_n
-k_m)z}}{A_n(z)A^*_m(z)}+\mathcal{O}\left(N^{-1}\right)\label{eq2.2}
\end{eqnarray}
with $z=x-y$.  In order to obtain Eq.~\eqref{eq2.2}  we used
\begin{eqnarray}
\int\limits_{0}^{L}dx A_j(x)A^*_l(x)e^{\imath (k_j-k_l) x}\label{eq2.3}
&=&A_j(1)A^*_l(1)\frac{e^{\imath (k_j-k_l) L} -1}{\imath(k_j -k_l)}\nonumber\\
&=&L\Big((k_j - \Lambda)^2 +c^2)\Big)\delta_{jl}+\mathcal{O}(N^{-1}).
\end{eqnarray}
Taking into account the normalization condition
\begin{eqnarray}
\int\limits_0^Ldx \int\limits_0^Ldy R(x,y)=N\label{eq2.4}
\end{eqnarray}
we obtain
\begin{eqnarray}
R(z)=\frac{\rho}{L}\left(1-\frac{1}{N\sum\limits_{ n=1}^{N+1}
\Big|A_n(1)\Big|^{-2}}\sum_{ n, m}^{N+1}\frac{e^{\imath(k_n
-k_m)z}}{A_n(z)A^*_m(z)}\right) + \mathcal{O}(N^{-1})\label{eq2.4a}.
\end{eqnarray}
In the thermodynamic limit the summations can be replaced by
integrations. Then by introducing the dimensionless quantities
\begin{eqnarray}
\hat{c}=c/k_F \qquad \hat{\Lambda}=\Lambda/k_F \qquad \hat{z}=k_F
z \qquad \hat{R}(\hat{z})=\frac{L}{\rho}R(z)\label{eq2.5}
\end{eqnarray}
the density-density correlation function acquires the form
\begin{eqnarray}
\hat{R}(\hat{z})=1-\frac{\hat{c}}{2v(1,\hat{c},\hat{K})}
\left|\int\limits_{-1}^{+1}dk \frac{\displaystyle
e^{\displaystyle\imath k\hat{z}}}{i(k -\hat{\Lambda})
+\hat{c}}\right|^2 +\mathcal{O}(N^{-1})\label{eq2.6}.
\end{eqnarray}
We discuss Eq.~\eqref{eq2.6} in three limits. First it is easily seen
that in the limit of infinite interaction strength the
density-density correlation function becomes identical with that
 of free Fermions i.e.
\begin{eqnarray}
\lim_{\hat{c}\rightarrow \infty}\hat{R}(\hat{z})= 1-
\left(\frac{\sin\hat{z}}{\hat{z}}\right)^2 \label{eq2.7}.
\end{eqnarray}
On the other hand the second term in Eq.~\eqref{eq2.6} vanishes
like $\hat{c}$ as $\hat{c} \rightarrow 0 $ and hence it follows that
\begin{eqnarray}
\lim_{\hat{c}\rightarrow 0}\hat{R}(\hat{z})= 1 \label{eq2.8}.
\end{eqnarray}
Finally we consider the behavior of the density-density
correlation function for $\hat{z} \to 0$. This is the density of
sea--Fermions at the position of the distinguishable particle. For
hardcore Bosons the local correlations have been studied in
Ref.~\cite{gan03}. According to Eqs.~\eqref{eq2.7} and
\eqref{eq2.8} we expect a transition from $\hat{R}(0)=1$ to
$\hat{R}(0)=0$ as $\hat{c}$ increases from zero to $+\infty$. The
integral in Eq.~\eqref{eq2.6} can be evaluated and the resulting
expression reads
\begin{eqnarray}
\hat{R}(0)=1-\frac{\hat{c}}{2v(1,\hat{c},\hat{K})}\left|\ln
\left(\frac{1+\imath(1-\hat{\Lambda})/ \hat{c}}
{1-\imath(1+\hat{\Lambda})/\hat{c}}\right)\right|^2\label{eq2.9}.
\end{eqnarray}
For $\hat{K}=0$ i.e. $\hat{\Lambda}=0$ this simplifies to Eq.~\eqref{r0c0}. In
Fig.~\ref{DD-corr3D} we show the plot of $\hat{R}(\hat{z})$ as
function of the distance $\hat{z}$ and the interaction strength
$\hat{c}$.

While for small distances density-density correlation function
decays monotonically from one to zero as $\hat{c}$ increases this
behavior changes at large distances and  $\hat{R}(\hat{z})$
exhibits a local minimum as function of $\hat{c}$. The
corresponding plot for the ground state is qualitatively similar
to that of Fig.~\ref{DD-corr3D}. Quantitatively they differ  in the form
of the  "edge" which marks the transition between monotonic and
non-monotonic behavior in $\hat{c}$. Compared with the ground
state this edge  becomes more and more smeared out as $\hat{K}$
increases.

\section{Summary \& Conclusion}\label{summary}

For a particle interacting via a $\delta$--potential with a
Fermi--sea, we calculated expectation values as well as the extra
 particle's equal time Green's functions and the density-density
correlation function in dependence of its momentum. The
many--body wave function could be expressed as a determinant. This
allowed us to obtain a compact expression for the Green's function
even for finite interaction strength.

For infinite interaction strength the thermodynamic limit can be
performed and the Green's function is a solution of Painlev{\'e}
V, with a short distance boundary condition, which depends on the
particle's momentum. Using the connection formulae for
Painlev{\'e} \RM{5} we found an algebraic decay for large distances
as $x^{-\beta}$ with $\beta=(\hat{K}^2+1)/2$, where $\hat{K}=K/k_{\rm F}$ is the
extra particle's momentum on the scale of the Fermi--momentum. The
particle undergoes a transition from a free Fermion ($\hat{K}=1$)
to a hardcore Boson ($\hat{K}=0$) as its momentum moves from the
edge to the core of the Fermi--sea.

Up to now we were unable to perform the thermodynamic limit of the
Green's function for finite interaction strength.
However with the determinantal structure of the Green's function
presented here, an analytical way to perform this 
limit even for finite interaction strength is not out of reach.

Yet our
numerical simulations together with the small distance analysis
seem to indicate that the interaction strength does not
change the Green's function  quantitatively but rather
renormalizes the parameters.
We conjecture that the relation of the Green's function to
solutions of Painlev{\'e} equations holds beyond the hardcore
limit. Such a relation indeed exists for hardcore Bosons and was
employed  \cite{cre81,mcc86b} to calculate the first order
corrections in $1/c$ to the hardcore result (see Eq.~\eqref{asymptotics}).

\begin{acknowledgements}
We acknowledge fruitful discussion with B.~Gutkin, A.M.~Lunde and
V.~Osipov. For helpful comments we thank V.~Leiss. CH acknowledges
financial support by Studienstifutung des deutschen Volkes. 
HK acknowleges financial support by the CSIC through the JAE program.
\end{acknowledgements}

\appendix

\section{Representation as T{\"o}plitz Determinant}\label{appA}

For finite interaction strength the determinant in
Eq.~\eqref{greendecom1} representing the Green's function is not a
T{\"o}plitz determinant due to the non constant diagonal entries.
However, in the limit $c\to \infty$  the Green's function has a
representation as T{\"o}plitz determinant. To reveal it we write
the quantities $g_{jl}$  in Eq.~\eqref{eq3.1.6} as
\begin{eqnarray}
\lim_{c\rightarrow \infty} g_{jl}=Lc^2 e^{-\imath(n_j
+n_l+1 +2\arctan(\lambda)/\pi)t}\left( (\lambda^2 +1) \delta_{jl}
-2(1 +\imath \lambda) \frac{\sin\big(n_{jl}t\big)}{\pi
n_{jl}}\right)  . \label{eq3.2.2}
\end{eqnarray}
Here we use the notations
\begin{eqnarray}
t=\frac{\pi}{L}(y-y')\geq 0\quad, \quad n_{jl}=j-l\label{eq3.2.3} \
\end{eqnarray}
and furthermore assume that the quantum number $n_j$ are given by
the set in Eq.~\eqref{ms} Now the diagonal entries are constant, i.~e.~
independent of $k_j$. For the  linear combination of the
quantities $g_{nm}$ as it appears in Eq.~\eqref{greendecom} this
yields
\begin{eqnarray}
\lim_{c\rightarrow \infty}\Big(g_{jl}- g_{jl+1}-g_{j+1l} +g_{j+1 l+1}\Big)
= Lc^2 e^{-\imath(n_j +n_l+1 +2\arctan(\lambda)/\pi)t}\gamma_{jl}(t,\lambda),\label{eq3.2.3a}
\end{eqnarray}
where we have defined
\begin{eqnarray}
\gamma_{jl}(t,\lambda)&=& (\lambda^2+1)\Big(2 \cos (t)
 \delta_{jl} -\delta_{jl+1}-\delta_{jl-1}\Big)- \frac{4}{\pi}
 (\imath\lambda +1)\frac{\sin (n_{jl}|t|) \cos( t)}{n_{jl}}\nonumber\\
&+&\frac{2}{\pi}(\imath\lambda +1)\left(\frac{\sin ((n_{jl}-1)|t|)}{n_{jl}-1}
+\frac{\sin ((n_{jl}+1)|t|)}{n_{jl}+1}\right)\label{eq3.2.4}.
\end{eqnarray}
The full Green's function (see Eq.~\eqref{greendecom}) then reads
\begin{eqnarray}
\lim_{c\rightarrow \infty}G(t)\propto e^{2\imath \arctan(\lambda)t/\pi}
 \det\left[\gamma_{jl}(t,\lambda)\right]_{j,l=1,\ldots,N} \label{eq3.2.5}.
\end{eqnarray}
Employing the  normalization condition $G(0)=1/L$   this evaluates
to
\begin{eqnarray}
\lim_{c\rightarrow \infty}G(t)=\frac{\displaystyle e^{2\imath \arctan(\lambda)t/\pi}}
{L(N+1)(\lambda^2 +1)^N} \det\left[\gamma_{jl}(t,\lambda)\right]_{j,l=1,\ldots,N}.
 \label{eq3.2.7}
\end{eqnarray}
Correspondingly the interaction part $G_{\rm I}(t)$ reads
\begin{eqnarray}
\lim_{c\rightarrow \infty}G_{\rm I}(t)=\frac{e^{-2\imath
N\arctan(\lambda) t/\pi}}{(N+1)(\lambda^2
+1)^N}\det\left[\gamma_{jl}(t,\lambda)\right]_{j,l=1,\ldots,N}.
\label{eq3.2.7-1}
\end{eqnarray}

For $\lambda=0$ Eq.~\eqref{eq3.2.7} is equivalent to the
representation of the Green's function for hardcore Bosons as T\"oplitz-determinant
\cite{len64,for02}. If on the other hand $\lambda\rightarrow
\infty$ we obtain from Eq.~\eqref{eq3.2.7}
\begin{eqnarray}
\lim_{\lambda \to \infty}\lim_{c\rightarrow \infty}G(t)=
\frac{\displaystyle e^{\imath t}}{L(N+1)}\underbrace{\det\left[(2
\cos(t) \delta_{j,l}
-\delta_{j,l+1}-\delta_{j,l-1})\right]_{j,l=1,\ldots,N}}_{\mathcal{L}_N(t)}.\label{eq3.2.9}
\end{eqnarray}
Expanding the determinant in the equation above yields the
following recursion
\begin{eqnarray}
\mathcal{L}_N(t)=2\cos(t)\mathcal{L}_{N-1}(t) - \mathcal{L}_{N-2}(t)\label{eq3.2.6}.
\end{eqnarray}
With help of the relation $2\cos(x)\sin(x)= \sin(x-y) +\sin(x+y)$
we see that its solution is given by
\begin{eqnarray}
\mathcal{L}_N(t)=\frac{\sin((N+1)t)}{\sin(t)}\label{eq3.2.10}.
\end{eqnarray}
Consequently Eq.~\eqref{eq3.2.9} acquires the form
\begin{eqnarray}
\lim_{\lambda \to \infty}\lim_{c\rightarrow \infty}G(t)=
\frac{\displaystyle e^{\imath
t}}{L(N+1)}\frac{\sin((N+1)t)}{\sin(t)}\label{eq3.2.11}.
\end{eqnarray}
Analogously one obtains for the interaction part
\begin{eqnarray}
\lim_{\lambda \to \infty}\lim_{c\rightarrow \infty}G_{\rm
I}(t)=\frac{e^{-\imath N t}}{N+1}
\frac{\sin((N+1)t)}{\sin(t)}\label{eq3.2.12a}.
\end{eqnarray}
Eqs.~\eqref{eq3.2.11} and \eqref{eq3.2.12a} correspond to the free
Fermion result.


\end{document}